\documentclass[apj]{emulateapj}

\usepackage{amsmath}        

\newcommand{\degree}{$^{\circ}$}

\newcommand{\perbeam}{\,beam$^{-1}$}

\shorttitle{VLA/{\it Chandra} observations of SS\,433}
\shortauthors{Miller-Jones et al.}

\begin{document}


\title{Coupled radio and X-ray emission and evidence for discrete
  ejecta in the jets of SS\,433}


\author{J. C. A. Miller-Jones\altaffilmark{1,2}}
\affil{NRAO Headquarters, 520 Edgemont Road, Charlottesville,
 Virginia, 22903}
\email{jmiller@nrao.edu}

\author{S. Migliari}
\affil{Center for Astrophysics and Space Sciences, University of
 California San Diego, 9500 Gilman Dr., La Jolla, CA 92093}

\author{R. P. Fender\altaffilmark{2}}
\affil{School of Physics and Astronomy, University of Southampton,
Highfield, Southampton, SO17 1BJ, UK}

\author{T. W. J. Thompson}
\affil{Center for Astrophysics and Space Sciences, University of
 California San Diego, 9500 Gilman Dr., La Jolla, CA 92093}

\author{M. van der Klis}
\affil{Astronomical Institute 'Anton Pannekoek', University of Amsterdam,
 Kruislaan  403, 1098 SJ, Amsterdam, The Netherlands}

\and

\author{M. M\'endez\altaffilmark{2}}
\affil{Kapteyn Astronomical Institute, Groningen University, 9700 AV,
 Groningen, The Netherlands}


\altaffiltext{1}{Jansky Fellow, National Radio Astronomy Observatory}
\altaffiltext{2}{Astronomical Institute 'Anton Pannekoek', University
 of Amsterdam, Kruislaan  403, 1098 SJ, Amsterdam, The Netherlands}


\begin{abstract}
We present five epochs of simultaneous radio (VLA) and X-ray ({\it
  Chandra}) observations of SS\,433, to study the relation between the
  radio and X-ray emission in the arcsecond-scale jets of the source.
  We detected X-ray emission from the extended jets in only one of the
  five epochs of observation, indicating that the X-ray reheating
  mechanism is transient.  The reheating does not correlate with the
  total flux in the core or in the extended radio jets.  However, the
  radio emission in the X-ray reheating regions is enhanced when X-ray
  emission is present.  Deep images of the jets in linear polarization
  show that outside of the core, the magnetic field in the jets is
  aligned parallel to the local velocity vector, strengthening the
  case for the jets to be composed of discrete bullets rather than
  being continuous flux tubes.  We also observed anomalous regions of
  polarized emission well away from the kinematic trace, confirming
  the large-scale anisotropy of the magnetic field in the ambient medium
  surrounding the jets.
\end{abstract}





\section{Introduction}
\label{sec:intro}
SS\,433 is one of the most persistent, stable, and well-studied
sources of relativistic jets in the Galaxy.  The system is a high-mass
X-ray binary at a distance of $5.5\pm0.2$\,kpc (Lockman, Blundell \&
Goss, 2007), accreting at a very high rate.  The consequent high
kinetic energy output in the jets has deformed the surrounding W\,50
nebula \citep{Beg80}.

Precession of the jets in SS\,433 is observed in both the
Doppler-shifted X-ray \citep{Wat86} and optical \citep{Mar79} emission
lines, at distances of $\sim 10^{11}$ and $\sim 10^{15}$\,cm
respectively from the centre of the system.  Radio observations
further confirm the precession scenario, with the radio jets tracing
out a corkscrew pattern formed by the projection onto the plane of the
sky of a sequence of ballistically-moving knots ejected from a jet
precessing about a fixed axis \citep{Hje81}.  The jet precession is
well fit by the kinematic model \citep{Abe79,Mil79,Fab79}, in which
antiparallel jets precess about an axis inclined at 78.83\degree\ to
the line of sight with a period of 162.5\,d, a precession cone opening
angle of 19.85\degree, and a jet speed of $0.2602c$ \citep{Sti02}.
The velocity of the jets has remained very stable since the original
identification of the Doppler-shifted emission lines in the source
spectrum \citep{Mar79}, although there is evidence for small
deviations from the mean velocity, of rms $0.014c$ \citep{Blu05}.

The polarization properties of the radio jets were studied in detail
by \citet{Sti04}.  They found evidence for a depolarization region
surrounding the core, outside of which the jets showed linear
polarization of up to 20 per cent, with a magnetic field aligned
parallel to the kinematic locus of the jets.  \citet{Rob07} also
examined the polarization properties of the jets, confirming the
depolarization region and the alignment of the magnetic field with the
local jet direction within 1\,arcsec of the core.  They also noted the
presence of anomalous highly-polarized emission off the kinematic
trace, indicating that the magnetic fields in the ambient medium
should be highly ordered on scales of thousands of AU, possibly due to
interactions with the jet and subsequent shearing of the existing
field.  \citet{Sti04} interpreted the alignment of the magnetic field
with the local jet direction as evidence that the jet was a continuous
plasma tube, rather than being composed of discrete radio-emitting
plasmons.

Arcsecond-scale X-ray jets were first observed in SS\,433 by Marshall,
Canizares \& Schulz (2002), and Migliari, Fender \& M\'endez (2002)
detected highly-ionised, Doppler-shifted iron lines at distances of
$10^{17}$\,cm from the binary core, implying reheating of baryonic
matter to temperatures in excess of $10^7$\,K downstream in the jets.
The X-ray jets do not have a static, long-term structure, but vary on
timescales as short as days, if not hours.  There is no correlation
between the X-ray emission and the precession phase of the jets
\citep{Mig05}.  The emission was initially explained as internal
shocks arising from velocity variations in the ejecta energising the
flow downstream in the jets \citep{Mig02}.  \citet{Sti04} proposed an
alternative explanation for the X-ray reheating, invoking the model of
Heavens, Ballard \& Kirk (1990), whereby sideways ram pressure on the
plasma tube could give rise to shocks which energised the plasma.
However, \citet{Mig05} observed rapid variability, whereby two knots
on different parts of the precession trace brightened sequentially,
which they ascribed to energisation by a faster underlying flow of
$\sim0.5c$.  Such unseen, highly-relativistic flows have recently been
inferred to exist in the Galactic neutron star X-ray binaries Scorpius
X-1 \citep{Fom01} and Circinus X-1 \citep{Fen04}.  An underlying flow
propagating outwards at high speed energises the mildly-relativistic
radio-emitting regions further downstream in the flow, lighting them
up via shock heating.

\section{Observations and data reduction}
\label{sec:obs}

In order to test the hypothesis of a fast, unseen flow energising the
flow downstream in the jets, we proposed for a sequence of four
simultaneous 10-ks observations with {\it Chandra} and the Very Large
Array (VLA), to track the evolution of the thermal and non-thermal
components of the emitting regions.  The High Resolution Camera (HRC)
on board {\it Chandra} was used, to maximise the spatial
($<0.5$\,arcsec) and temporal (16\,$\mu$s) resolution, thus avoiding
pile-up and enabling us to identify and track the motions of individual
jet knots.  The observing frequency in the radio band was chosen in
order to best match the spatial resolution of {\it Chandra}.

The observations were spaced over the course of 8 days in 2006
February, and the radio and X-ray properties are summarised in
Tables~\ref{tab:obs} and \ref{tab:xobs}.  We also re-analysed archival
radio and X-ray data from July 2003, in order to investigate for the
first time the balance of thermal and non-thermal emission from the
arcsecond-scale X-ray jets.

\subsection{Radio}
In both the observations of 2003 July and those of 2006 February,
we observed SS\,433 at 4.86\,GHz with the VLA in its A-configuration.
Table \ref{tab:obs} shows the observation dates and durations for each
epoch.  The observing bandwidth was 50\,MHz.  The primary calibrator
was 3C\,286, and the secondary calibrator was J\,1922+1530.  The flux
density scale used was that derived at the VLA in 1999 as implemented
in the 2004 December 31 version of {\sc aips}.  In the observations of
2003 July, the parallactic angle coverage of J\,1922+1530 was
sufficient to calibrate the polarization leakage terms, whereas in
February 2006 the calibrator 3C\,84 was used for this part of the
calibration (except on 2006 February 19, when it was not observed and
no polarization calibration was possible).  In all cases, 3C\,286 was
used to calibrate the absolute polarization position angle.  External
gain and polarization calibration were automated using {\sc
ParselTongue}, a Python interface to {\sc aips}.  Self-calibration and
imaging were then carried out manually, following standard procedures
within {\sc aips}.

\begin{deluxetable*}{cccccccccc}
\setlength{\tabcolsep}{0.015in} 
\tablecolumns{10}
\tablewidth{0pc}
\tabletypesize{\scriptsize}
\tablecaption{Radio properties of the jets of SS\,433.  Precessional phase
  is defined according to the convention of \citet{Sti02}, whereby
  $\phi=0,1,2,...$ corresponds to the eastern jet having maximum
  redshift. Column 3 gives the flux density of the core at each
  epoch, column 4 the total integrated flux density of the core and extended
  jets taken together, and columns 5 and 6 the total integrated flux
  density, $I$,
  within the eastern and western regions shown in  Fig.~\ref{fig:jul_overlay}.
  Columns 7 and 8 give the total integrated polarized flux density, $P$, for
  those same eastern and western regions, and columns 9 and 10 give
  the average fractional polarizations, $<F>=P/I$,
  within those regions.  The quoted uncertainties
  on the flux densities are just the statistical uncertainties due to the rms
  noise in the images.  There is a further systematic
  uncertainty in the absolute flux calibration, believed to be of
  order 1--2 per cent.\label{tab:obs}}
\tablehead{
\colhead{} & \colhead{} & \multicolumn{4}{c}{Flux densities} & \multicolumn{2}{c}{Polarized flux densities} & \multicolumn{2}{c}{Fractional polarization}\\
\colhead{Date} & \colhead{Phase} & \colhead{Core} & \colhead{Jet (integrated)} & \colhead{East} & \colhead{West} & \colhead{East}
& \colhead{West} & \colhead{East} & \colhead{West}\\
\colhead{(MJD)} & \colhead{} & \colhead{(mJy bm$^{-1}$)} & \colhead{mJy} & \colhead{mJy} &
\colhead{mJy} & \colhead{mJy} & \colhead{mJy} & \colhead{} & \colhead{}}
\startdata
$52831.29\pm0.22$ & 0.94 & $307.2\pm0.02$ & $474.5\pm0.7$ &
$14.2\pm0.3$ & $23.1\pm0.3$ & $3.0\pm0.1$ & $2.8\pm0.1$ &
$0.21\pm0.01$ & $0.12\pm0.01$\\
$53785.80\pm0.08$ & 0.82 & $525.6\pm0.06$ & $667.1\pm0.5$ &
$7.8\pm0.6$ & $9.9\pm0.6$ & - & - & - & -\\
$53787.76\pm0.08$ & 0.83 & $472.1\pm0.03$ & $623.7\pm1.0$ &
$6.9\pm0.4$ & $8.7\pm0.4$ & $1.3\pm0.3$ & $1.2\pm0.3$ & $0.19\pm0.04$ & $0.14\pm0.03$\\
$53790.83\pm0.07$ & 0.85 & $484.8\pm0.03$ & $636.6\pm1.3$ &
$6.4\pm0.4$ & $9.2\pm0.4$ & $0.9\pm0.2$ &
$0.9\pm0.2$ & $0.14\pm0.04$ & $0.10\pm0.03$\\
$53793.68\pm0.07$ & 0.87 & $539.0\pm0.03$ & $716.3\pm1.1$ &
$7.1\pm0.4$ & $9.3\pm0.4$ & $1.2\pm0.3$ &
$1.0\pm0.3$ & $0.17\pm0.05$ & $0.11\pm0.04$\\
\enddata
\end{deluxetable*}

\subsection{X-ray}

\subsubsection{ACIS}
\label{sec:acis}
\begin{figure}
\begin{center}
\plotone{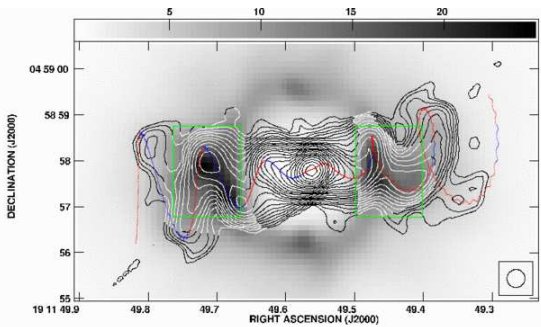}
\caption{{\it Chandra} ACIS-S X-ray greyscale image of the 60-ks
  observation of 2003 July with radio contours superposed.  The
  intensity wedge shows the ACIS-S count rates in counts\,s$^{-1}$.
  Total intensity contours are at levels of $\pm(\sqrt{2})^n$ times
  19.8\,$\mu$\,Jy\,bm$^{-1}$, where $n=4,5,6,...$.  The maximum flux
  density in the image is 307\,mJy\perbeam.  The restoring beam, shown
  in the lower right of the image, is $0.40\times0.39$\,arcsec$^2$ in
  PA $-25$\fdg5.  The calculated kinematic model precession trace,
  including nodding, has been overlaid.  Red and blueshifted ejecta
  are indicated by their respective colours.  The two green boxes show
  the X-ray extraction regions.}
\label{fig:jul_overlay}
\end{center}
\end{figure}

\begin{figure}
\begin{center}
\plotone{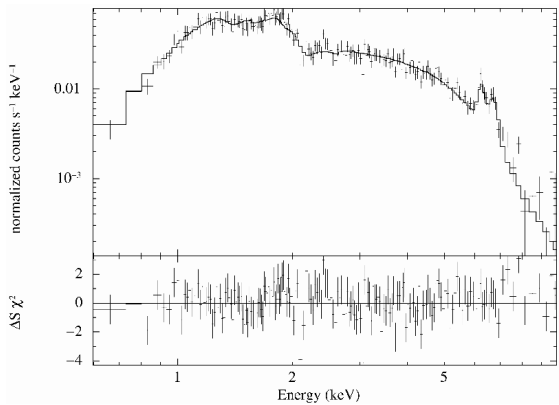}
\caption{Spectrum of the core during the {\it Chandra} ACIS-S
observation of 2003 July 10--11, extracted from the readout streak. The
normalization shown is not corrected for the effective exposure
time. The spectrum of the core is well fitted with a power-law model,
an absorption edge at 1.3 keV, and two iron emission lines at
$\sim6.2$ and $\sim6.7$\,keV (see Section \ref{sec:acis}).
\label{fig:core_spec} }
\end{center}
\end{figure}

The 60-ks ACIS-S observation of 2003 July 10--11 (MJD 52830--52831),
previously analysed by \citet{Mig05}, shows two X-ray extensions, east
and west from the heavily piled-up core (Fig.~\ref{fig:jul_overlay}).
We estimated the 0.5--10~keV spectrum of the core by analyzing the
photons in the readout streak, which should be unaffected by pile-up.
We extracted the background-subtracted spectrum in a rectangular
region with $280\times 20$ pixels covering the readout streak, and
corrected it with the effective exposure time for the readout streak
data (see Migliari et al.\ 2005 for details). We show the spectrum of
the core in Fig.~\ref{fig:core_spec}. The spectrum is well fitted with
a model consisting of an absorbed power law, with photon index $\sim
1.6$, two Gaussian emission lines at $\sim6.2$ and $\sim6.7$\,keV, and
an absorption edge around 1.3 keV (at a significance of $7.4\sigma$),
obtaining a reduced $\chi^2=1.21$ with 172 d.o.f.\ (Table
\ref{tab:xfits}). The core flux, i.e.\ the flux in the readout streak
spectrum corrected for the effective exposure time, is
$7.1\times10^{-10}$ erg\,cm$^{-2}$\,s$^{-1}$.

\begin{figure}
\begin{center}
\plotone{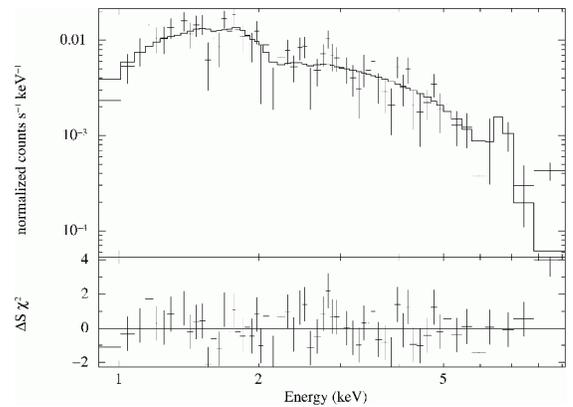}
\caption{Core-PSF subtracted spectrum of the eastern jet region during
the {\it Chandra} ACIS-S observation of 2003 July 10--11. The continuum is
well fitted with either a 10-keV bremsstrahlung continuum (shown here)
or a power-law component. The iron emission line at $\sim6.7$\,keV is
marginally significant (see Section \ref{sec:acis}).
\label{fig:east_spec} }
\end{center}
\end{figure}

To extract the spectra of the X-ray jets, we selected two regions of
size $1.46\times1.98$\,arcsec$^2$, east and west of the core, as shown
in Fig.~\ref{fig:jul_overlay}. We extracted the 0.5--10 keV energy
spectra in these two regions and subtracted the contribution of the
core due to the tail of the point spread function (PSF). To do so, we
created the PSF of a point source, using the PSF shape for photons at
roughly the average energy of the spectrum, i.e.\ 3\,keV, and with a
count rate normalized to that estimated from the analysis of the
readout streak. We chose the core-tail regions to subtract, such that
in the PSF image their counts were equal to those in the jet region.
In Fig.~\ref{fig:east_spec}, we show the core-tail subtracted spectrum
of the eastern jet, re-binned with a minimum of 40 counts per energy
channel.  The eastern jet spectrum is well fitted, with a reduced
$\chi^2$ of 1.12 (58 d.o.f.), using an absorbed bremsstrahlung model
with a temperature of $\sim10$ keV and a marginally-detected Gaussian
emission line around 6.7\,keV (the normalization with $1\sigma$ error
bars was
$4.7^{+2.1}_{-1.6}\times10^{-6}$\,photons\,cm$^{-2}$\,s$^{-1}$).
Given the broadening of the PSF at higher energies, we note that the
use of a 6.7-keV PSF reduces the line strength by about a factor of
two.  We obtained an equally good fit with an absorbed power-law model
($\Gamma=1.7\pm0.3$), and cannot distinguish between the
bremsstrahlung and power-law models with the current data.  We note
that the emission line around 6.2\,keV present in the spectrum of the
core (Fig.~\ref{fig:core_spec}) disappears completely.  Freezing the
6.7-keV line parameters and adding an extra line to the model, with
the energy and width set to 6.2 and 0.07\,keV respectively (as found
in the core), gave an upper limit to the normalization of
$6.7\times10^{-7}$\,photons\,cm$^{-2}$\,s$^{-1}$.  The corresponding
upper limit to the flux in the line is
$<6.7\times10^{-15}$\,erg\,cm$^{-2}$\,s$^{-1}$.  The 2--10 keV flux in
the eastern jet region is $3.5\times10^{-13}$
erg\,cm$^{-2}$\,s$^{-1}$. The best-fitting parameters of this model
are shown in Table \ref{tab:xfits}.  The flux estimate for the western
X-ray jet region, being very close to the core, is more difficult to
constrain due to the high pile-up. Its core-tail subtracted energy
spectrum is well fitted with a simple absorbed bremsstrahlung model
with a temperature of $\sim3$ keV and has a 2-10\,keV flux upper limit
(98 per cent confidence) of $2.6\times10^{-13}$
erg\,cm$^{-2}$\,s$^{-1}$ (see Table \ref{tab:xfits}).

\begin{deluxetable}{cccc}
\tablecolumns{4}
\tablewidth{0pc}
\tabletypesize{\small}
\tablecaption{Best-fit parameters of the 0.5--10\,keV energy spectra in the
readout streak, east and west regions of the 2003 July {\it
Chandra}/ACIS-S obervation.\label{tab:xfits}} 
\tablehead{\colhead{} & \colhead{Core} & \colhead{East} & \colhead{West}
}
\startdata
$N_{\rm H}$ & 0.79$\pm0.02$&0.9$\pm0.1$ & 1.9$\pm0.6$\\
 & & & \\
$\Gamma$  & $1.48\pm0.02$& $1.7\pm0.3$ & --\\
$N_{\rm pl}$ & 4.0$\pm0.2$& $0.87^{+0.45}_{-0.25}$ &-- \\
 & & & \\
$kT_{\rm br}$ & $26\pm4$ & 10.0$^{+5.2}_{-2.5}$& 3.1$\pm1.8$\\
$N_{\rm br}$  & $57\pm1$ &  9.7$\pm0.9$& $<15$ \\
 & & & \\
$E_{\rm g1}$ & $6.23^{+0.07}_{-0.03}$ & 6.2 & -- \\
$\sigma_1$ &$0.08\pm0.02$  & 0.07 & -- \\
$N_{g1}$ & 12$\pm2$ & $<0.67$  &  --\\
 & & & \\
$E_{\rm g2}$ & $6.68\pm0.03$ &   6.7$\pm0.1$& -- \\
$\sigma_2$ & $0.17^{+0.41}_{-0.04}$ &  $<0.3$&  --\\
$N_{\rm g2}$ & $19\pm3$ & $4.7^{+3.4}_{-2.7}$ &  --\\
& & & \\
$E_{\rm edge}$ & $1.35\pm0.02$ & -- & -- \\
$\tau$ &  0.37$\pm0.05$& -- &  --\\
&  &  &  \\
$F_X$ & $7.1\times10^{-10}$ & $3.5\times10^{-13}$ & $<2.6\times10^{-13}$\\
&  &  &  \\
$\chi^2/{\rm d.o.f.}$& 208/172 & 65/58 & 61/40 \\
\enddata
\tablecomments{The parameters are: equivalent hydrogen
column density $N_{\rm H}$ (in units of $10^{22}$\,cm$^{-2}$), power
law normalization $N_{\rm pl}$ (in units of
$10^{-4}$\,photons\,keV$^{-1}$\,cm$^{-2}$\,s$^{-1}$ at 1 keV) and
photon index $\Gamma$, bremsstrahlung temperature $kT_{\rm br}$ (in
keV) and normalization $N_{\rm br}$ (in units of $(3.02 \times 10^{-20}/(4
\pi D^2)) \int n_{\rm e} n_{\rm i} dV $, where $D$ is the source
distance (cm), and $n_{\rm e}$ and
$n_{\rm i}$ are the electron and ion densities (cm$^{-3}$)), Gaussian
emission line peak energy $E_{\rm g}$ (in keV), Gaussian line width
$\sigma_{\rm g}$ (in keV) and normalization N$_{\rm g}$ (in
$10^{-6}$\,photons cm$^{-2}$ s$^{-1}$ in the line), absorption edge
energy threshold $E_{\rm edge}$ (in keV) and maximum absorption factor at
threshold $\tau$ (in keV), and total X-ray flux $F_{X}$ (in
erg\,cm$^{-2}$\,s$^{-1}$).  In the readout streak the normalizations of
the model components have not been corrected for the effective exposure
time.  Note that, due to the proximity of the heavily piled-up core,
the spectrum of the west region is not well constrained.}
\end{deluxetable}

\subsubsection{HRC}
\begin{figure*}
\begin{center}
\plotone{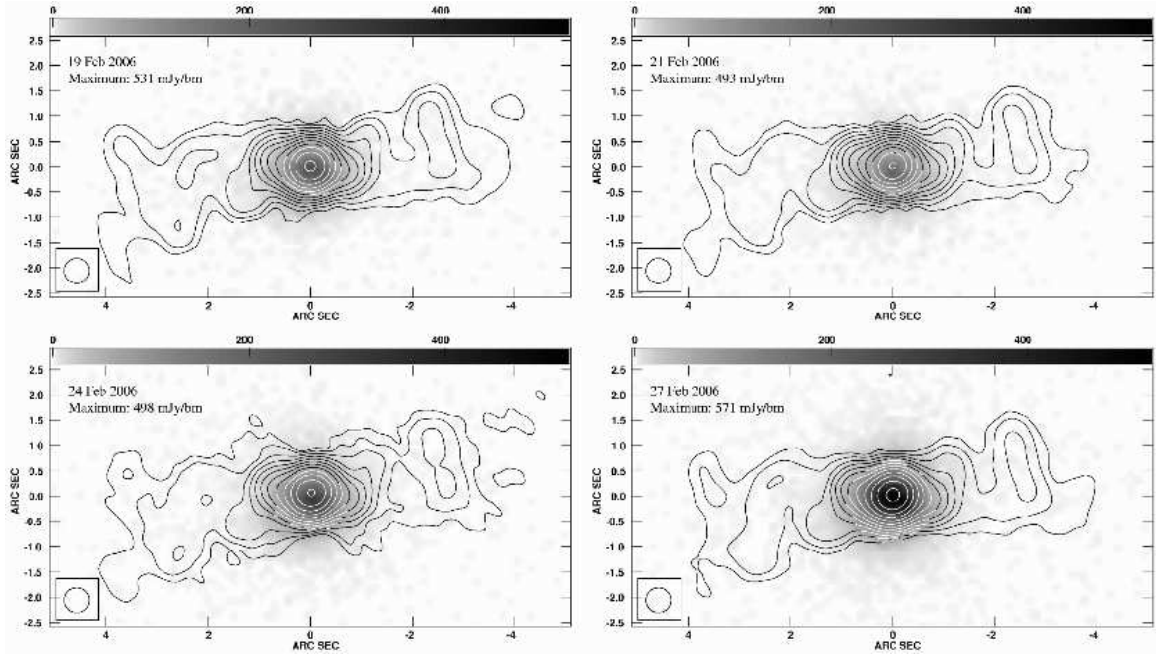}
\caption{{\it Chandra} HRC greyscale images from 2006 February, with
  radio contours overlaid.  The intensity wedges (all set to the same
  minimum and maximum) show the total counts from each HRC
  observation.  Total intensity contours are at levels of $\pm2^n$
  times a representative rms noise level of 46\,$\mu$Jy\,bm$^{-1}$,
  where $n=5,6,7,...$.  The maximum total intensity in each image is
  given in the top left-hand corner.  All images have been restored
  with the same beamsize of $0.49\times0.49$\,arcsec$^2$ (shown in the
  lower right corner of each image).  There is no evidence for any
  X-ray extension along the jet direction in any case.  The X-ray core
  gets brighter over the course of the last three observations.
\label{fig:xr_montage} }
\end{center}
\end{figure*}

For the four HRC observations of 2006 February
(Fig.~\ref{fig:xr_montage}), we extracted the counts in the same east
and west regions used in the 2003 July ACIS observation.  To estimate
the contribution of the core, we created the normalized PSF of each
image and extracted the counts in those same east and west regions.
The PSF subtracted count rates of the east and west regions for the
four HRC observations are shown in Table \ref{tab:xobs}.  There was no
clear evidence in the HRC observations for jet-like extended emission
as seen in the ACIS image, although the count rate was not consistent
with zero.  The images (Fig.~\ref{fig:xr_montage}) showed slightly
enhanced diffuse emission in the E-W direction, possibly due to
scattering of the core emission from the denser material along the jet
direction.  Other than this, the source was consistent with being an
unresolved, variable point source.  Since the peak sensitivity of the
HRC is in the energy range 0.2--2\,keV, the ACIS images of 2003 July
were remade, using only the 0.5--2\,keV energy range.  The
arcsecond-scale X-ray jets were still present in this low-energy data
from 2003 July, ruling out the possibility that the extension was only
present in hard X-rays ($>2$\,keV).

To facilitate comparison of the measured X-ray count rates with
earlier observations, we converted the measured HRC 0.1--10\,keV count
rates to those that would have been measured by HETGS (0.8--8\,keV)
and ACIS-S (0.5-10\,keV).  For this we used PIMMS with a model
consisting of a power law with photon index $\Gamma = 1.66$ (the mean
of the parameters derived for the HETGS observations by
\citet{Mig05}), affected by interstellar absorption with $N_{\rm H} =
1\times10^{22}$\,cm$^{-2}$.

\begin{deluxetable*}{cc|ccc|ccc|ccc}
\setlength{\tabcolsep}{0.03in} 
\tablecolumns{11}
\tablewidth{0pc}
\tabletypesize{\scriptsize}
\tablecaption{Net count rates (after PSF subtraction) of the HRC
  observations in the extraction regions of the core, eastern and
  western jets of SS\,433.  The 0.1--10\,keV HRC count rates have been
  converted to the 0.8--8\,keV HETGS and 0.5--10\,keV ACIS-S count
  rates, for comparison with the count rates derived by
  \citet{Mig05}.\label{tab:xobs}}
\tablehead{\colhead{} & \colhead{} & \multicolumn{3}{|c|}{Core} & \multicolumn{3}{|c|}{Eastern Jet} & \multicolumn{3}{|c}{Western jet}\\
\colhead{Date} & \colhead{Instrument} & \colhead{HRC} &
\colhead{HETGS} & \colhead{ACIS} & \colhead{HRC} & \colhead{HETGS} &
\colhead{ACIS} & \colhead{HRC} & \colhead{HETGS} & \colhead{ACIS}\\
\colhead{(MJD)} & \colhead{} & \colhead{s$^{-1}$)} & \colhead{(s$^{-1}$)} & \colhead{(s$^{-1}$)} & \colhead{($10^{-3}$ s$^{-1}$)} & \colhead{($10^{-3}$ s$^{-1}$)} & \colhead{($10^{-3}$ s$^{-1}$)} & \colhead{($10^{-3}$ s$^{-1}$)} & \colhead{($10^{-3}$ s$^{-1}$)} & \colhead{($10^{-3}$ s$^{-1}$)}
}
\startdata
51444 & HETGS & & 1.442 & & & $3.6\pm1.4$ & & & & \\
51876 & HETGS & & 0.557 & & & $18.3\pm1.3$ & & & & \\
51984 & HETGS & & 1.876 & & & $24.6\pm2.1$ & & & & \\
52037 & HETGS & & 0.363 & & & $23.4\pm1.3$ & & & & \\
52039 & HETGS & & 0.234 & & & $25.4\pm1.4$ & & & & \\
52041 & HETGS & & 0.846 & & & $23.8\pm1.7$ & & & & \\
51722 & ACIS-S & & & 2.3 & & & $43.7\pm2.5$ & & & \\
52830 & ACIS-S & & & 8.8 & & & $27.8\pm1.1$ & & & $<1.2$\\
53785 & HRC & $0.753\pm0.007$ & 0.610 & 3.1 & $3.72\pm0.84$ & 3.016 &
15.44 & $8.07\pm1.27$ & 6.542 & 33.49\\
53787 & HRC & $0.568\pm0.006$ & 0.460 & 2.4 & $4.47\pm0.76$ & 3.624 &
18.55 & $8.54\pm1.08$ & 6.923 & 35.44\\
53790 & HRC & $0.997\pm0.008$ & 0.808 & 4.1 & $4.38\pm0.90$ & 3.551 &
18.18 & $7.31\pm1.29$ & 5.926 & 30.34\\
53793 & HRC & $1.816\pm0.011$ & 1.472 & 7.5 & $4.80\pm1.22$ & 3.891 &
19.92 & $11.60\pm1.57$ & 9.403 & 48.14\\
\enddata
\end{deluxetable*}

\subsubsection{The nature of the iron line}
At the time of the 2003 July observation, the precessional phase was
0.94 (in the convention that phase 0, 1, 2,... correspond to the point
at which the eastern jet is maximally redshifted).  This would predict
redshifts of 0.001 and 0.073 in the western and eastern jets
respectively.  This would make the two lines seen in the core spectrum
(Table \ref{tab:xfits}) consistent with the red and blueshifted Fe XXV
1s2p-1s$^2$ transition (rest energy 6.68\,keV); the redshifts derived
from the precessional phase would predict 6.68 and 6.23\,keV
respectively.  Since the western jet has redshift 0.001, this line
would be blended with emission from the same line in the stationary
core, explaining both the higher normalization and broader line
profile of the 6.68\,keV line.  We note that this transition (rest
wavelength 1.855\,\AA) was also the most prominent line seen in the
HETGS observations of \citet{Mar02}, albeit at a different orbital
phase and hence with different redshifts in the jets, lending further
credence to our interpretation of the line as highly-ionised iron.

The iron line in the extended emission in the eastern jet is less easy
to identify.  The X-rays are spatially extended, so from the position
of the extraction box (Fig.~\ref{fig:jul_overlay}), the redshift could
lie anywhere between -0.11 and 0.08 (the maximum possible blueshift
and redshift for the eastern jet), giving a rest energy between 5.96
and 7.24\,keV.  However, Fig.~\ref{fig:jul_overlay} shows that the
X-ray emission is strongest along the blueshifted part of the trace,
and the location of the X-ray bright spots are both close to the
transition between blueshift and redshift.  If the line emission
traces the continuum, we can rule out any redshift, so the rest
frequency of the line would be $\leq6.7$\,keV.  The kinematic model
calculations for the positions of the X-ray hotspots suggest that any
blueshift is $\geq -0.04$.  We can thus rule out the 7.06\,keV Fe XXV
k$\beta$ line seen by \citet{Mig02}, although both the neutral Fe-K
line (6.40\,keV), and the Fe XXV line seen in the core spectrum are
plausible candidates.  With such a marginal line detection however,
we caution against over-interpreting this result.

\section{The radio images}

\subsection{The kinematic model trace}
\begin{figure*}
\begin{center}
\epsscale{0.8}
\plotone{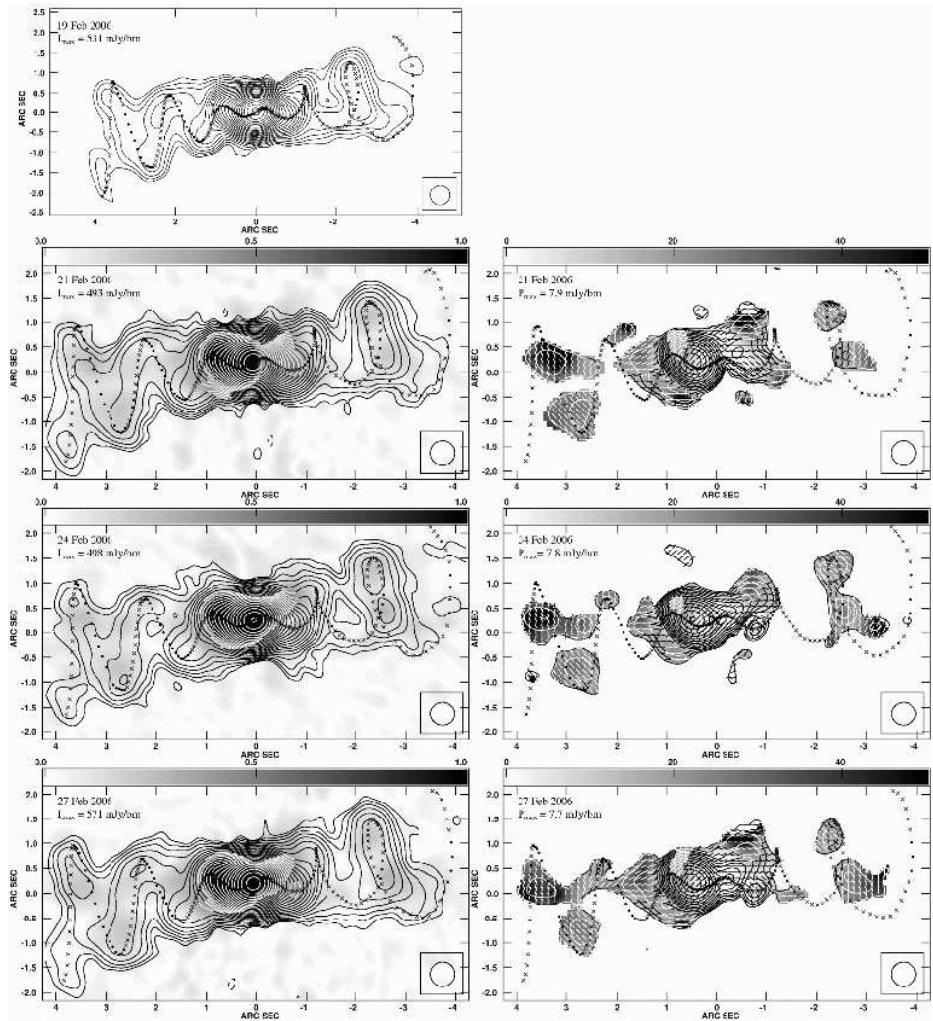}
\caption{Radio images of SS\,433 in 2006 February.  Left column shows
  total intensity contours superposed on polarized intensity
  greyscale.  Total intensity contours are at levels of
  $\pm(\sqrt{2})^n$, with $n=5,6,7,...$ times the noise level in the
  individual images (47.1, 35.4, 47.3 and 32.7\,$\mu$Jy\,bm$^{-1}$
  respectively), and the greyscale intensity wedge gives the polarized
  intensity in mJy.  Right column shows polarized intensity contours
  superposed on fractional polarization greyscale ($F=P/I$), with
  polarization vectors showing the EVPA.  Polarization vectors are of
  uniform length and have been rotated by $-25.9$\degree\ to correct
  for Faraday rotation.  The strength of polarization is given by the
  contours, which are at levels of $\pm(\sqrt{2})^n$ times the noise
  level in the polarized images (27, 24 and 26\,$\mu$\,Jy\,bm$^{-1}$
  respectively), where $n=5,6,7,...$.  The greyscale intensity wedge
  is in per cent.  The date and maximum intensity of each image are
  given in the top left-hand corner.  All images have been restored
  with the same beamsize of $0.49\times0.49$\,arcsec$^2$ (shown in the
  lower right corner of each image).  The calculated precession trace
  for each date, including the nodding parameters given by
  \citet{Sti02}, has been overlaid.  Red and blueshifted ejecta are
  indicated by crosses and filled circles respectively.
\label{fig:fpol} }
\end{center}
\end{figure*}

\begin{figure}
\begin{center}
\epsscale{1}
\plotone{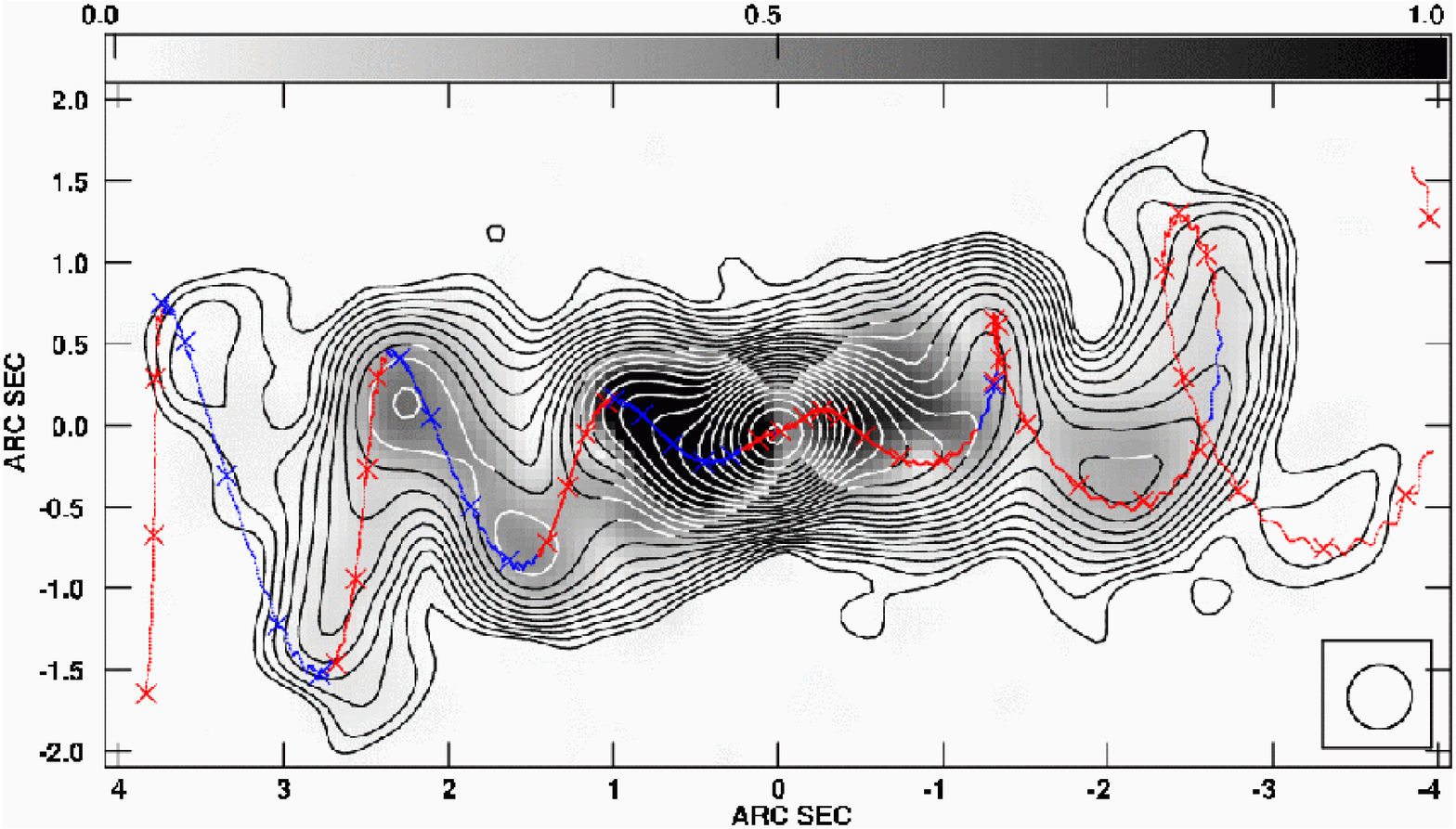}
 \caption{Total intensity contours of 2003 July superposed on linear
  polarization greyscale.  Total intensity contours are at levels of
  $\pm(\sqrt{2})^n$ times the noise level of
  19.8\,$\mu$\,Jy\,bm$^{-1}$, where $n=5,6,7,...$.  The greyscale
  intensity wedge is in mJy\perbeam.  The maximum flux density in the image
  is 307\,mJy\perbeam.  The restoring beam, shown in the lower right
  of the image, is $0.40\times0.39$\,arcsec$^2$ in PA $-25$\fdg5.  The
  calculated precession trace for each date, including the nodding
  parameters given by \citet{Sti02}, has been overlaid.  Red and
  blueshifted ejecta are indicated by their respective colours.  The
  crosses along the trace mark ejecta separated by 20\,d.}
\label{fig:jul_ifpol}
\end{center}
\end{figure}

To calculate the kinematic model predictions, we used the published
parameters of \citet{Sti02}, but with the velocity and distance
derived by \citet{Blu04}, $0.2647c$ and 5.5\,kpc respectively.  We
note that the phase reference time quoted by \citet{Sti02} is in TJD
rather than MJD, so it should read MJD\,48615.5, for the point at
which the eastern jet is maximally redshifted.  With these parameters,
the predicted trace appeared to overlay the data well, as seen in
Figs.~\ref{fig:jul_overlay}, \ref{fig:fpol}, \ref{fig:jul_ifpol}, and
\ref{fig:jul_fpol}.  We saw no need to invoke a quarter-period phase
delay, a deceleration or a slow jet, as suggested by \citet{Sti04}.

The left column of Fig.~\ref{fig:fpol} shows the total intensity
contours and radio polarization measurements (where polarization
calibration was available) for all four epochs of observation during
2006 February, with the fitted kinematic model trace for each epoch
overlaid.  There is no significant evidence for the successive
brightening of knots along the kinematic trace suggested by
\citet{Mig05}.  This suggests that any underlying relativistic flow
within the jets is not persistent.  The rise in the core flux density
over the course of the last three epochs of observation (Table
\ref{tab:obs}), taken together with the rising X-ray count rate of the
core, suggests that a new phase of activity and jet heating might have
been beginning at this time.

\subsection{Polarization}
\citet{Sti04} studied the polarization properties of the
arcsecond-scale radio jets of SS\,433 and found a depolarization
region surrounding the core of extent $\sim0.5$\,arcsec at 5\,GHz.
They derived a mean value of the rotation measure (RM) towards the
source of $119\pm16$\,rad\,m$^{-2}$, which increased to
$\sim300$\,rad\,m$^{-2}$ close to the radio core.

The right column of Fig.~\ref{fig:fpol} shows the derived fractional
polarization and electric vector position angles (EVPAs) for each of
the epochs of observation in 2006 February where polarization data
were available.  The polarization vectors have been rotated clockwise
by 25.9\degree, to compensate for the rotation measure of
$119\pm6$\,rad\,m$^{-2}$ found by \citet{Sti04}.  The images have all
been convolved with the same circular restoring beam, to facilitate
comparisons.  The fractional polarization (greyscale), is shown only
for the regions where both the total intensity and the linear
polarization were more significant than $5\sigma$.  Fractional
polarizations of up to 30 per cent are seen along the jets.

\begin{figure}
\begin{center}
\epsscale{1}
\plotone{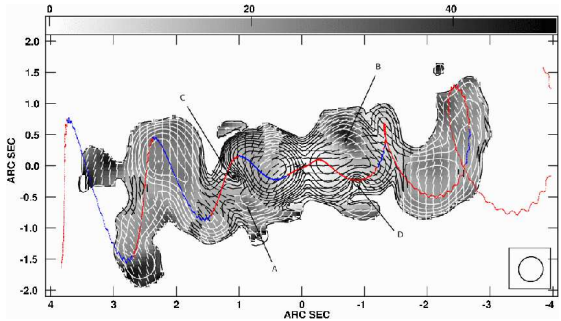}
 \caption{Linear polarization contours of 2003 July superposed on
  fractional polarization greyscale with EVPA vectors overlaid.  The
  polarization vectors have been rotated by $-25.9$\degree\ to correct
  for the Faraday rotation due to the global RM of 119\,rad\,m$^{-2}$.
  The magnetic field is perpendicular to the EVPAs.  The polarization
  vectors are all of uniform length.  The strength of polarization is
  given by the contours, which are at levels of $\pm(\sqrt{2})^n$
  times the noise level in the polarized image of
  10.3\,$\mu$\,Jy\,bm$^{-1}$, where $n=5,6,7,...$.  The greyscale
  intensity wedge is in per cent.  The maximum polarized flux density in
  the image is 4.5\,mJy\perbeam.  The restoring beam, shown in the
  lower right of the image, is $0.40\times0.39$\,arcsec$^2$ in PA
  $-21$\fdg3.  The calculated precession trace for each date,
  including the nodding parameters given by \citet{Sti02}, has been
  overlaid.  Red and blueshifted ejecta are indicated by their
  respective colours.  The central cross marks the position of the
  radio core (maximum total intensity).}
\label{fig:jul_fpol}
\end{center}
\end{figure}

The linear polarization image of 2003 July 11, shown in
Fig.~\ref{fig:jul_fpol}, confirms the depolarization of the core.
Moving outwards, the fractional polarization then levels off at 20--30
per cent along the rest of the jet until the signal drops below the
noise level.  The high degree of polarization confirms the radio
emission to be of synchrotron origin, ruling out thermal free-free
emission as a viable emission mechanism.

\subsubsection{Magnetic field orientation}
\label{sec:bfield}
\begin{figure}
\begin{center}
\plotone{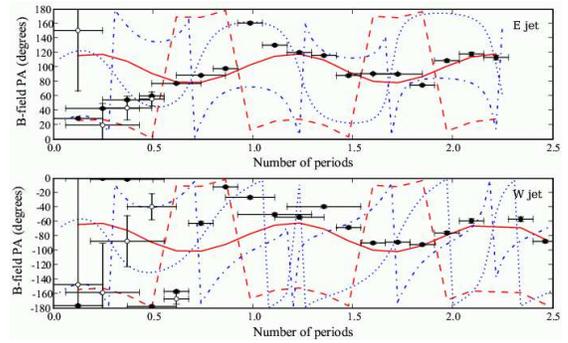}
 \caption{Plots of the variation of the magnetic field direction with
  position away from the core along the kinematic model trace.  The
  magnetic field direction has in all cases been assumed to be
  perpendicular to the Faraday rotation-corrected EVPA.  Filled
  circles are for Faraday derotation with the global RM of
  119\,rad\,m$^{-2}$ and open circles use the RMs derived from a fit
  to the data of \citet{Rob07}.  Red solid and dashed lines show the
  predictions of the kinematic model for the directions parallel and
  perpendicular to the local velocity vector respectively.  Blue dotted and
  dot-dashed lines show the directions along and perpendicular to the
  local jet direction (the projection of the kinematic model trace on
  the plane of the sky) respectively.  All field directions have a
  180\degree\ ambiguity, and the model traces have been wrapped by
  180\degree\ where necessary.  The top plot shows the approaching
  (eastern) jet and the bottom plot is for the western (receding jet).
  The measured points far enough away from the core correlate well in
  both jets with the solid line tracing the local velocity vector (the
  direction of motion of the ballistically-moving ejecta).}
\label{fig:bfield}
\end{center}
\end{figure}

With our deep image of 2003 July 11 (Figs.~\ref{fig:jul_ifpol} and
\ref{fig:jul_fpol}), whose rms of 19.8\,$\mu$Jy\,bm$^{-1}$ is a factor
3.5 lower than the 5-GHz observation of \citet{Sti04}, we can more
accurately measure the magnetic field direction along the jet.
Fig.~\ref{fig:bfield} shows the magnetic field directions inferred
from the assumption that the field direction is perpendicular to the
EVPA.

From about half a period downstream of the core (corresponding to a
distance of $\sim0.7$\,arcsec) in the approaching jet and one full
period in the receding western jet (where the rotation measure is
greater), the implied magnetic field direction is extremely
well-aligned parallel to the local velocity vector.  This is in
contrast to the work of \citet{Sti04}, who suggested that the field
was aligned along the kinematic model locus out to $\sim1.5$
precession periods ($\sim2.5$\,arcsec) from the core.  Closer to the
central binary system, this correlation clearly breaks down.  However,
since the EVPAs have only been corrected for the global foreground
Faraday rotation of 119\,rad\,m$^{-2}$, the true EVPA close to the
core will be smaller than the measured values.  This effect is
noticeable further out in the receding (western) jet than in the
approaching (eastern) jet, consistent with the results of
\citet{Rob07}, whose figure 11 demonstrates clearly that the fitted
rotation measure in the western jet exceeds that in the eastern jet
out to $\sim0.9$\,arcsec from the core.  However, using their fitted
RMs does not bring the implied magnetic field direction back to
alignment with the local jet velocity (see the white points in
Fig.~\ref{fig:bfield}).  The source is optically thin down to
$\sim30$\,mas from the central binary \citep{Par99}, so the deviation
from the correlation close to the core cannot be explained by the
transition from the case where the magnetic field is aligned
perpendicular to the EVPA to one where it is parallel, as occurs on
moving from an optically thin to thick regime.  We note that
\citet{Rob07} found that the magnetic field between $\sim0.4$ and
1\,arcsec of the core appeared to follow the kinematic model trace,
becoming perpendicular to the trace in the very inner regions (although
with such a small fractional polarization in the central regions, this
result was not deemed to be secure).  Higher-resolution observations
at multiple frequencies are required to determine whether the magnetic
field orientation really changes along the jets.

The images of 2006 February are less deep, such that we cannot
reliably measure the polarization and magnetic field orientation all
along the kinematic trace.  However, the observed EVPA orientations
in the outer parts of the jets (Fig.~\ref{fig:fpol}) also appear to be
perpendicular to the direction radially outwards from the core,
suggesting that the alignment of the magnetic field with the local
velocity vector is persistent.

\subsubsection{Anomalous polarized emission}
The polarized intensity does not follow the total intensity (and hence
the kinematic trace) perfectly.  Regions A and B in
Fig.~\ref{fig:jul_fpol} show high linear and fractional polarization
and are well off the kinematic model trace.  Regions C and D, on the
contrary, lie directly on the kinematic model trace, and show very low
linear and fractional polarization, which can not be explained by beam
depolarization due to the kinematic model covering a full rotation in
those regions.  Region C corresponds to the discrepant point in the
top plot of Fig.~\ref{fig:bfield}, at 0.98 periods, so the low
fractional polarization seems to be related to the anomaly in magnetic
field direction.

We note that regions A and B correspond well to knots 1 and 2
identified in figure 13 of \citet{Sti04}.  These knots, and also the
anomalous knots observed by \citet{Spe84}, appear at the point in the
precession trace where the jet makes its first significant bend.
Similar features are also seen in the polarization images of the 2006
February observations (Fig.~\ref{fig:fpol}).  From
Fig.~\ref{fig:bfield}, this ($\sim0.4$ periods downstream from the core; see
Fig.~\ref{fig:bfield}) is just after the point at which there is
a local maximum in the angle between the local velocity vector and the
local jet direction (solid and dotted lines).  The Faraday-derotated
EVPAs in these regions are aligned pointing back towards the core,
implying that the magnetic field is perpendicular to the velocity, if
these regions have a velocity away from the central binary.  This
could be caused by compression of the magnetic field at a shock front
moving outwards \citep[e.g.][]{Sti04,Hea90}, although the origin of
such a shock remains unclear.  \citet{Rob07} also found
highly-polarized off-jet emission on the inner side of the first bend
in the jet, which they ascribed to large-scale shearing motions in the
ambient medium close to the jets.  Either way, such structures are
clearly not unusual in this system.

\section{Comparison of radio and X-ray data}
\label{sec:comparison}
No extended X-ray emission was seen in the observations of 2006
February, and there was no evidence in the radio images for a series
of knots along the precession trace brightening sequentially.  We are
thus unable to verify the scenario of a faster-moving shock lighting
up individual knots as it passes \citep{Mig05}.  If such a fast,
unseen flow exists, it was not present during the course of these
observations, and further monitoring is required when extended X-ray
jets are present, to verify this scenario.

The only observation in which both radio and X-ray emission were
observed from the arcsecond-scale jets was that of 2003 July.  Before
a comparison could be made, the two sets of images were reprojected to
the same geometry using the {\sc aips} task OHGEO.  While the SIN
projection is common in radio aperture synthesis, the {\it Chandra}
data used the TAN projection, necessitating a reprojection of the
radio image before creating the overlay shown in
Fig.~\ref{fig:jul_overlay} (see \citet{Gre83} for further details).
This Figure shows that the X-ray emitting regions lie on the kinematic
model trace, and correlate well with the radio emission from the jet.

\subsection{The broadband spectrum}
\begin{figure}
\begin{center}
\plotone{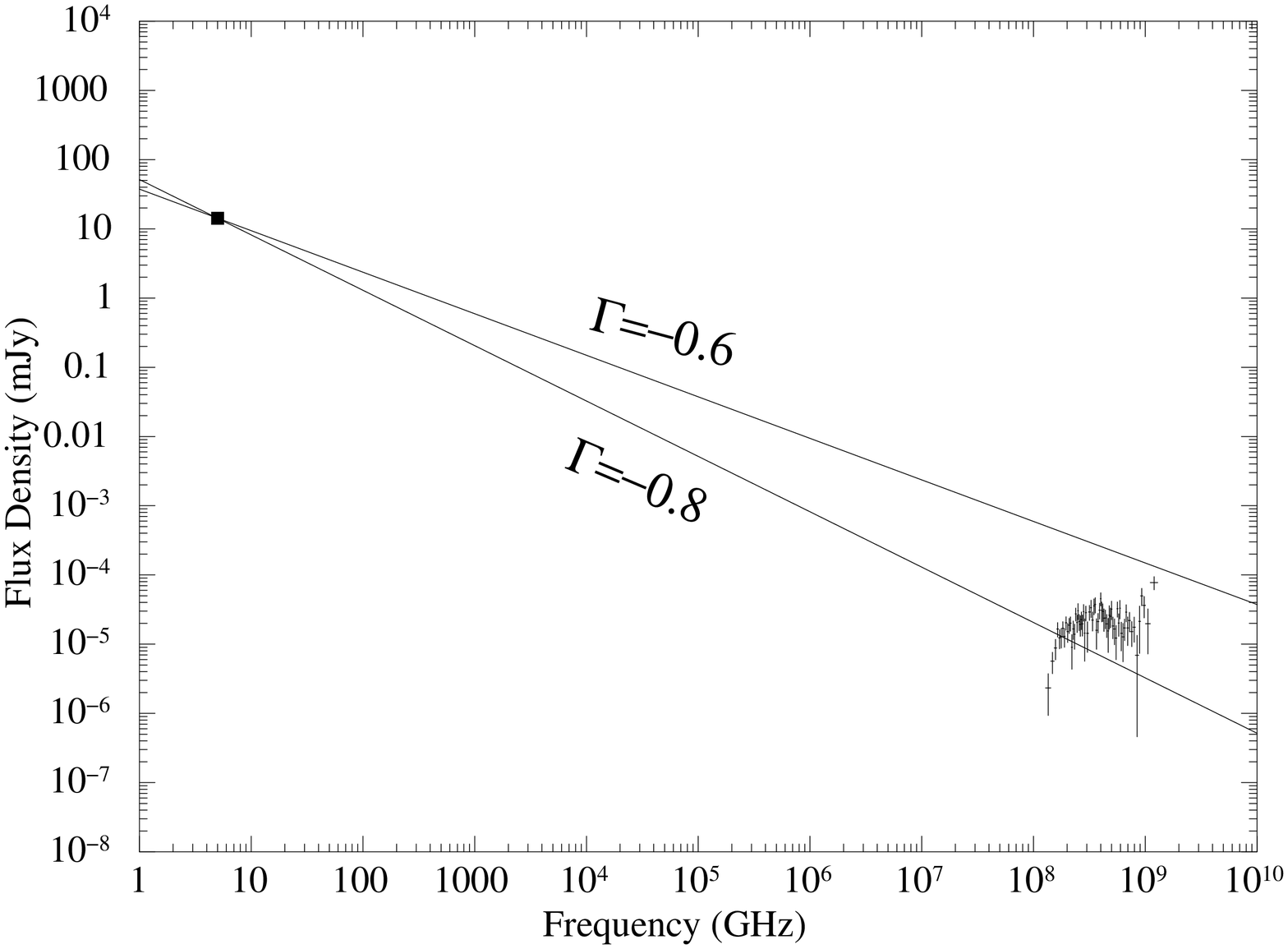}
\caption{Broadband radio to X-ray spectrum of the eastern X-ray jet
  region for 2003 July.  Power law spectra of spectral index $-0.6$
  and $-0.8$ have been overlaid, demonstrating the need for accurate
  radio spectral indices to help constrain the nature of the X-ray
  emission.
\label{fig:broadband_spectrum} }
\end{center}
\end{figure}

Since a fit to the X-ray spectrum cannot significantly distinguish
between power-law and bremsstrahlung emission, we cannot determine the
emission mechanism from the X-ray data alone.  A synchrotron spectrum
with a spectral index of $-0.7$ to $-0.8$
($S_{\nu}\propto\nu^{\alpha}$) seems to be consistent with the overall
radio to X-ray spectrum (Fig.~\ref{fig:broadband_spectrum}), and with
the power-law fit to the X-ray data.  Our single-frequency radio
observations do not allow us to empirically determine the spectral
index.  Previous authors \citep{Ver93,Sti04} have measured spectral
indices of between $-0.6$ and $-0.8$ in the extended jets outside the
core region, which would be consistent with a single spectrum running
from the radio to the X-ray regime.  However, if the X-ray emission
has the same origin as that detected by \citet{Mig02}, the
highly-ionized iron lines seen in that observation would argue for a
high temperature and bremsstrahlung X-ray emission, implying a hybrid
thermal/non-thermal plasma in the jets.

\subsection{A radio/X-ray correlation}
We measured the total radio flux density enclosed in the regions
selected for the X-ray data analysis (see Fig.~\ref{fig:jul_overlay})
in the images of all 5 epochs, and the results are presented in Table
\ref{tab:obs}.  In 2003 July, when the X-ray reheating was observed,
the radio flux was enhanced by a factor $\sim 2$ in the regions
containing the X-ray emission, despite both the core emission and the
integrated radio flux density in the jets being significantly less
bright.  This demonstrates that the radio emission is indeed related
to the X-ray emission, and that there is some connection between the
populations of emitting particles.  The average fractional
polarization does not seem to change significantly, implying that
while the radio emission mechanism is synchrotron, the reheating does
not change the degree of ordering of the magnetic field in the jets.

\citet{Mig05} presented evidence for variability of the
arcsecond-scale X-ray jets in SS\,433, on timescales of days and
possibly hours.  The X-ray jets are not persistent structures, as
expected from the $\sim2$\,d lifetime of optically-emitting bullets
close to the core \citep{Ver93b}.  From our non-detections in 2006
February, we can put a lower limit on the duty cycle of any underlying
shocks of $\sim10$\,d.  This is further evidence for the transient
nature of the X-ray jets and the reheating process that creates them.

\subsection{Energetics considerations}
The X-ray and radio fluxes in the selected X-ray emitting region in
the eastern jet in 2003 July can be used to calculate the mass of particles
required to generate the emission.  Assuming a source distance of
5.5\,kpc and a cylindrical emission region of length 1.46\,arcsec and
radius $1.98/2=0.99$\,arcsec, we derive an emitting volume of
$2.5\times10^{45}$\,m$^3$.

\subsubsection{X-rays}
Assuming the X-ray emission to be thermal, we use the bremsstrahlung fit to
the X-ray spectrum, which gave a temperature of 10.0\,keV, and a 2--10\,keV
X-ray flux of $3.5\times10^{-13}$\,erg\,cm$^{-2}$\,s$^{-1}$.  The flux
from thermal bremsstrahlung can be expressed as \citep{Lon94}
\begin{equation}
F_{\rm X} = \frac{1.435\times10^{-40}Z^2T^{1/2}gNN_{\rm e}Vf}{4\pi
  d^2}\quad {\rm W\,m^{-2}},
\end{equation}
where $Z$ is the mean atomic number of the atoms, $T$ is the
temperature, $g$ is the Gaunt factor, and $N$ and $N_{\rm e}$ are the
ion and electron number densities.  Assuming one proton for every
electron, then $N=N_{\rm e}$.  Thus we find a baryonic mass associated
with the bremsstrahlung X-ray emission of
\begin{equation}
M_{\rm brem} = m_{\rm p}\left(\frac{4\pi d^2 F_{\rm X}
  Vf}{1.435\times10^{-40}Z^2 T^{1/2} g}\right)^{1/2}\quad {\rm kg},
\end{equation}
where $m_{\rm p}$ is the proton mass.  We do not know a priori the
filling factor $f$ of the emitting region.  However, the predicted
mass transfer rates are hyper-Eddington (King, Taam \& Begelman 2000),
so we can assume the accretion rate to be close to Eddington.
K\"ording, Fender \& Migliari (2006) found that a constant fraction of
the liberated accretion power is injected into the jet.  The kinetic
power of the jet in SS\,433, $(\gamma-1)\dot{M}c^2$, should therefore
be of order $L_{\rm Edd}$.  The kinetic energy flux through the region
where the X-ray counts are estimated is
\begin{equation}
L_{\rm kin} = (\gamma-1)M_{\rm brem}c^2 \frac{vf}{l} = 10^{-3}L_{\rm
  Edd},
\end{equation}
where $v$ is the jet speed, $c$ is the speed of light, $\gamma$ is the
Lorentz factor, and $l$ is the length of the emitting
region along the direction of the jets.  Rearranging and solving for
the filling factor, using the fitted X-ray flux and temperature, a
canonical Gaunt factor of 1.2, and assuming $Z=1$ for a baryonic
component consisting mainly of hydrogen, we find $f=3\times10^{-3}$.
To create the observed bremsstrahlung emission, this then gives an
associated mass in baryons of $1.7\times10^{-5}M_{\odot}$.

\subsubsection{Radio}
Since the radio emission appears to be enhanced in the X-ray reheating
region, we can attempt to estimate the mass required to emit the
observed radio flux.  Classic synchrotron minimum energy arguments,
assuming that the electron energy distribution, $N(E)=\kappa E^{-p}dE$,
extends from $E_{\rm min}=m_{\rm e}c^2$ up to $E_{\rm max}\gg m_{\rm
e}c^2$, where $m_{\rm e}$ is the electron mass, give a total energy in
relativistic particles of
\begin{multline}
W_{\rm part} = \left(\frac{4\pi d^2 S_{\nu} \nu^{(p-1)/2} \left(m_{\rm
    e}c^2\right)^{2-p}}{2.344\times10^{-25}(1.253\times10^{37})^{(p-1)/2}}\right)^{4/(5+p)}\\
\times\left[(p-2)a(p)\right]^{-4/(5+p)}\left(\frac{\mu_0(1+p)}{2Vf}\right)^{-\frac{1+p}{5+p}},
\end{multline}
where $d$ is the source distance, $\mu_0$ is the permeability of free
space, and $a(p)$ is a slowly varying factor of $p$, equal to $0.45$
for $p=2.2$.  We can integrate the electron energy distribution to
find the total number of relativistic electrons responsible for the
observed emission, via
\begin{equation}
N = \int^{E_{\rm max}}_{E_{\rm min}} \kappa E^{-p}\,dE \sim
\frac{\kappa (m_{\rm e} c^2)^{1-p}}{(p-1)}.
\end{equation}
By integrating to find the total energy in relativistic particles, we
can solve for $\kappa$.  We find
\begin{equation}
\kappa = \frac{W_{\rm part}(p-2)}{\eta Vf (m_{\rm e}c^2)^{2-p}}.
\end{equation}

Putting this all together, we can express the total mass of baryons
associated with the synchrotron emission detected in the radio band as
\begin{equation}
M_{\rm synch} = VfNm_{\rm p} = \frac{W_{\rm part} m_{\rm p} (p-2)}{\eta
  m_{\rm e}c^2 (p-1)}.
\end{equation}
Assuming one proton for every electron, a canonical spectral index
$\alpha=-0.6$ ($p=2.2$), a volume $V=2.5\times10^{45}$\,m$^{3}$ and a
filling factor of $f=3\times10^{-3}$ as derived above, we find a
baryonic mass of $7.1\times10^{-10}M_{\odot}$ associated with the
radio emission, variable within a factor $\sim 5$ for the range of
spectral indices considered, i.e.\ $2.0<p<2.6$.  The radio-emitting
particles form a very small fraction of the total population in the
jets.  Most of the energy in the jets is dark (i.e\ kinetic, rather
than radiative), in the absence of X-ray reheating.

\section{The magnetic field direction and implications for the nature
  of the jets}

There has been extensive debate in the literature as to the continuous
or discrete nature of the jets in SS\,433.  The presence of multiple
peaks in the Doppler-shifted H$\alpha$ line profiles of the source was
taken as evidence for the discrete nature of the jets \citep{Ver93b}.
\citet{Cha02} discussed potential mechanisms for producing bullet-like
ejections, and \citet{Cha03} presented multiwavelength variability as
evidence for such discrete bullets.  However, the continuous core
wings seen in VLBI images, the lack of structure on long baselines,
and the tails of the bright radio knots which appeared to trace the
nodding motion, argued for a more continuous flow, albeit with
intermittent brightness enhancements \citep{Ver93}.  The latter
authors postulated the existence of a two-phase jet, with dense, cool
optical-emitting clumps existing in a hot, continuous flow.  The
alignment of the magnetic field with the local velocity vector found
by \citet{Sti04} was taken as further evidence for continuous jets.

We observed the magnetic field to be preferentially aligned with the
ballistic velocity of the jet knots once they are sufficiently far
outside the core to be unaffected by the enhanced Faraday rotation
present close to the central binary.  The discrepancy of our results
with those of \citet{Sti04} is likely to be due to the higher
signal-to-noise of our observations and the consequently increased
number of precession periods out to which we can probe the magnetic
field direction.  Since the correlation only becomes clear $\gtrsim 1$
period out from the core (Fig.~\ref{fig:bfield}), several precession
cycles must be measured in order to detect the true correlation.
Closer in, the increasing RM causes a rotation of the EVPAs and
implied magnetic field directions, such that without an accurate RM
map for the central regions, accurate measurements of the position
angle cannot be made.  However, \citet{Rob07} measured the magnetic
field direction at 14.94\,GHz, where the Faraday rotation, even for an
RM of 596\,rad\,m$^{-2}$ (their highest measured value) is less than
15\degree, which should not be sufficient to destroy the measured
correlation.  They found that the field was preferentially aligned
along the kinematic model trace (the jet ridge line) between $\sim0.4$
and 1\,arcsec from the core.  This suggests that the alignment of the
magnetic field could change on moving beyond the core region, either
because of contamination from the surroundings or the emission of the
disc wind \citep{Par99,Blu01}, or possibly reflecting a real change in
the field orientation, perhaps due to shocks compressing the field at
some distance from the core.

\citet{Sti04} derived the expected EVPA orientation for a longitudinal
magnetic field along the plasma tube in the case of a smooth,
continuous jet, finding it to be perpendicular to the apparent
direction of the flux tube on the plane of the sky.  Hence the
inferred magnetic field direction would lie along the locus of the
kinematic model trace, in contrast to the alignment with the ballistic
velocity of the jet components (i.e.\ radially outward from the core)
seen in our data beyond 1\,arcsec.  This argues against a smooth,
continuous jet, since the absence of a component of the field
perpendicular to the direction of ballistic motion implies that, to
first order, the field at each point along the jet is not influenced
by the direction of motion of the neighbouring points, such that each
point is independent.

\citet{Sti04} also discussed the expected magnetic field
configurations for a knotty jet composed of discrete components.  If
the ejecta compress the ISM through which they are moving, the ensuing
shocks at the front surface will create a component of the magnetic
field perpendicular to the velocity (e.g.\ Hughes, Aller \& Aller,
1989).  Alternatively, if the knots expand freely as they move through
well-defined channels, \citet{Sti04} suggested that entrainment or
stretching of the field lines would generate a component of the field
parallel to the velocity vector.  Due to the precession, such channels
would appear as an evacuated layer on the surface of a cone, which,
given the relatively long precession period of 162.5\,d, we consider
unlikely to remain empty for an entire precession cycle.  We therefore
consider alternative possible explanations for the observed field
direction.

A preferential field orientation does not necessarily imply true
unifomity of the magnetic field, since observations are only sensitive
to the degree of ordering of the field in the plane perpendicular to
the line of sight.  An essentially irregular field can be made
anisotropic via compression or shear effects, and depending on the
orientation of the line of sight with respect to the plane of the
shear or compression, can result in high degrees of observed
polarization and apparent ordering of the field \citep{Lai80}.
Velocity shear along the direction of motion and compression of the
field via lateral expansion of an ejected knot could both give rise to
an apparent ordering of the field parallel to the local velocity
vector, although this would come about in different ways.  Shear would
reduce the radial component of the field (perpendicular to the
velocity vector) while leaving the longitudinal (parallel to the jet
velocity) and toroidal components unchanged, whereas compression would
amplify the longitudinal and toroidal components, without affecting
the strength of the radial component.

The observed values of 20--30 per cent polarization along the jet
locus are far from the theoretical maximum of $3(p+1)/(3p+7)$ (70 per
cent for the canonical value of $p=2.2$), implying some depolarization
within the beam or along the line of sight.  The jets are unresolved,
so it is possible for the true geometry to comprise a number of
emitting structures with different field alignments, with the
emission-weighted average over the beam determining the observed
orientation.  Shock-compressed magnetic fields at the front surface of
the knots could still be present, diluting the observed percentage
polarization from a well-ordered field aligned parallel to the
velocity, generated by lateral expansion and shearing effects.  A
possible analogy is to the hotspots of FR II radio galaxies
\citep[e.g.][]{Dre87}.  A bowshock forms around the hotspot as it
propagates outwards, with a field orientation perpendicular to the
direction of motion at the head of the jet, and field lines swept back
around the edges of the radio lobe, giving the field a ``U''-shaped
geometry.  In our data, this entire structure is unresolved by the
VLA.  Close to the central source \citep[$<0.4$\,arcsec;][]{Rob07},
the shock at the front surface is brightest, giving an observed field
perpendicular to the velocity.  As lateral expansion and velocity
shear create a field well-aligned with the direction of motion, the
parallel component begins to dominate and the observed field
eventually becomes parallel to the velocity (i.e.\ the emission is
dominated by long tails at the two sides of the ``U'', illuminated by
the swept-back particles accelerated at the working surface).  The
corkscrew pattern observed for the SS\,433 jets would then be caused
by a superposition of many of these unresolved hotspot structures
propagating outwards at different angles due to the precession of the
jets.

Without resolving the emitting knots, it is impossible to derive the
true field configuration and verify this scenario, which, while
plausible, remains speculative.  Detailed modelling of the true
emissivity, magnetic field and velocity distributions across the
source \citep[as done for AGN by][]{Lai04} is impossible without
further constraints, and is beyond the scope of this paper.  The best
approach would be to make deep observations of the jets at high
frequencies in relatively extended array configurations, to attempt to
resolve the individual emitting knots well outside the core, and the
variation in magnetic field orientation across them.  Furthermore,
high frequency observations, where the data are less affected by
Faraday rotation, are warranted to probe the inner regions, to
ascertain whether the magnetic field direction remains correlated with
the local velocity close to the central binary.  Whatever the actual
field configuration, the observed correlation of the magnetic field
direction with the local velocity vector (Section \ref{sec:bfield})
implies that the jet is unlikely to be a continuous flux tube, and is
more likely to be composed of a series of discrete, unconnected,
radio-emitting bullets, as seen in VLBI images
\citep[e.g.][]{Mio03,Par99}.

\section{Conclusions}
With the deepest radio polarization image to date, we have measured
the magnetic field direction in the extended jets, finding it to be
aligned with the local velocity vector.  This suggests that the jets
are likely to be composed of discrete bullets, rather than being
continuous flux tubes.  We confirm the depolarization region
surrounding the core which was noted by \citet{Sti04}.  We also see
evidence for regions of anomalous, polarized emission away from the
kinematic model trace in images from five different epochs, seen also
in older observations from the literature.  Such anomalous emitting
structures appear to be the rule, rather than the exception in this
system.  The magnetic field in these anomalous regions is observed to
be aligned perpendicular to the velocity vector away from the core,
indicating possible shock compression of the field.  Alternatively,
shearing motions at the interface of the jet and ambient medium could
cause the observed large-scale ordering of the magnetic field.

For the first time, we have studied the correlation between the radio
and X-ray emission in the arcsecond-scale jets of SS\,433.  The radio
emission is enhanced in the presence of X-ray jets.  The
arcsecond-scale X-ray jets are transient, and their presence does not
correlate with the flux density of the core.  We saw no evidence for a
faster, underlying flow energising individual emitting knots.  If it
exists in this system, it was not active during these observations.  The
core shows evidence for highly-ionised iron lines, at the redshifts
predicted by the kinematic model.  We cannot confirm at any
significant level the existence of the iron line previously seen in
the spatially resolved X-ray jets, because of the very bright and
highly piled-up core which makes the measurement of the weak features
in the jet spectra, such as the lines, very difficult.  The measured
X-ray flux is consistent with a single spectral index of $-0.8$
running all the way from the radio band to the X-rays.  While we
cannot significantly distinguish between a bremsstrahlung and a
power-law origin for the X-rays in the extended jets, bremsstrahlung
seems more likely given the highly-ionised iron lines observed
previously in the jets, which would suggest that the X-rays are not
simply the tail of the radio synchrotron spectrum.

\acknowledgements
The authors would like to thank Alan Bridle for useful discussions.
JCAM-J is grateful for support from the National Radio Astronomy
Observatory, which is operated by Associated Universities, Inc., under
cooperative agreement with the National Science Foundation.  SM
acknowledges support under the grants GO6-7030X and NNG056E70G.  TT
acknowledges support under the grant NAS5-30720.  ParselTongue was
developed in the context of the ALBUS project, which has benefited
from research funding from the European Community's sixth Framework
Programme under RadioNet R113CT 2003 5058187.
{\it Facilities:} \facility{CXO (ACIS,HRC)}, \facility{VLA}

\bibliographystyle{mn2e}

\end{document}